\documentclass[twocolumn,aps,superscriptaddress]{revtex4-2}

\usepackage{graphicx}
\usepackage{dcolumn}
\usepackage{bm}
\usepackage{amsmath,amssymb,amsfonts}
\usepackage{color}
\usepackage{braket}
\usepackage{mathtools}

\begin{document}

\title{Toroidal Scattering and Nonreciprocal Transport by Magnetic Impurities}

\author{Hiroki Isobe}
\affiliation{Department of Applied Physics, University of Tokyo, Bunkyo, Tokyo, 113-8656 Japan}

\author{Naoto Nagaosa}
\affiliation{RIKEN Center for Emergent Matter Science (CEMS), Wako, Saitama 351-0198, Japan}
\affiliation{Department of Applied Physics, University of Tokyo, Bunkyo, Tokyo, 113-8656 Japan}

\begin{abstract}
We propose the second-order response of metals in an electric field induced by magnetic impurities which locally break inversion symmetry. 
The impurities with toroidal moments scatter conduction electrons in the presence of the spin-orbit coupling, leading to nonreciprocal response. 
This mechanism is ubiquitous when a magnetic impurity is placed off an inversion center such as an interstitial site and a surface of a two-dimensional system. 
\end{abstract}

\maketitle

A magnetic toroidal moment often refers to a spatial texture of magnetic moments, which breaks both inversion and time-reversal symmetries while the product remains intact \cite{Dubovik,Gorbatsevich,Schmid,Spaldin}.  
However, even a single magnetic moment can generate a finite toroidal moment when it is placed off an inversion center of the underlying crystal structure.  
To discuss a toroidal moment by a magnetic impurity in a metal, we consider the spin-orbit coupling and the exchange coupling between the 
conduction electrons and a magnetic impurity 
\begin{equation}
H_{\rm imp} = V(\bm{r}-\bm{r}_i) + \operatorname{Sym}[ \lambda_0 (\bm{\sigma} \times 
\nabla V(\bm{r}-\bm{r}_i) ) \cdot \bm{p} ]
- J \bm{\sigma} \cdot \bm{s}_i, 
\end{equation}
where $V(\bm{r}-\bm{r}_i)$ is the potential created by an the impurity at $\bm{r}_i$, $\bm{p}$ is the momentum operator of conduction electrons, and $\operatorname{Sym}[AB] = (AB+BA)/2$ is the symmetrizing operation.  

\begin{figure}[b]
\centering
\includegraphics[width=0.7\hsize]{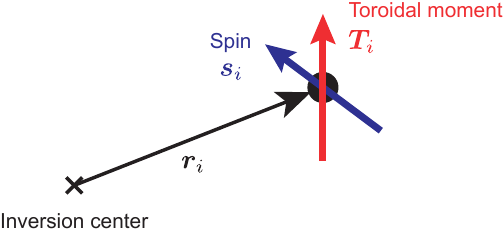}
\caption{Magnetic toroidal moment from a single magnetic element.    
}
\label{fig:toroidal}
\end{figure}

When the exchange coupling is much stronger than the spin-orbit coupling, we can replace the conduction electron spin $\bm{\sigma}$ in the spin-orbit term with the impurity spin $\bm{s}_i$ to obtain the effective Hamiltonian
\begin{equation}
H_i(\bm{r}) = V(\bm{r}-\bm{r}_i) + \operatorname{Sym}[ \lambda \bm{T}_i(\bm{r}- \bm{r}_i)  \cdot \bm{p} ], 
\label{eq:effective}
\end{equation}
which shows the coupling between the magnetic impurity spin $\bm{s}_i$ and the conduction electron spin $\bm{\sigma}$. 
Here, $\lambda$ is the spin-orbit coupling and the toroidal moment $\bm{T}_i(\bm{r}- \bm{r}_i)$ is given by 
\begin{equation}
\bm{T}_i(\bm{r}- \bm{r}_i) =  \bm{s}_i \times \nabla V(\bm{r}-\bm{r}_i) . 
\label{eq:T}
\end{equation}
When the impurity potential $V(\bm{r}-\bm{r}_i)$ depends only on $ |\bm{r}-\bm{r}_i|$, its gradient becomes $\nabla V(\bm{r}-\bm{r}_i) \propto \bm{r}-\bm{r}_i$ and the spatial average results in
\begin{equation}
\langle \bm{T}_i \rangle \propto  \bm{s}_i \times \bm{r}_i. 
\label{eq:T2}
\end{equation} 
We obtain the common expression $\bm{T}_i \propto \bm{r}_i \times \bm{s}_i$, where $\bm{r}_i$ is measured from the inversion center (Fig.~\ref{fig:toroidal}), and the total toroidal moment of a magnetic texture formed by a group of spins $\{ \bm{s}_i \}$ is given by $\bm{T}= \sum_i \bm{r}_i \times \bm{s}_i$. 
An important remark here is that a spin cluster is not needed for the discussion below. 
Each spatially-separated impurity spin individually contributes to the scattering of the conduction electrons.  

Since a toroidal moment breaks inversion symmetry, an external electric field $\bm{E}$ can induce second-order current response \cite{MA}  
\begin{equation}
j_a(\omega+\omega') = \chi_{abc}(\omega,\omega') E_b(\omega) E_c(\omega'), 
\end{equation}
where $\bm{j}$ is the current density and $\chi_{abc}$ is the second-order conductivity tensor.  
The summation over the repeated indices is implicit.  
We assume that the underlying material is nonmagnetic and its crystal structure preserves inversion symmetry since the host material in itself can be centrosymmetric for finite second-order response.  
We suppose the spin-unpolarized energy band with the dispersion $\epsilon_{\bm{k}} = \bm{k}^2/(2m)$.  A plane-wave state is an energy eigenstate, denoted by $\ket{\bm{k}} = e^{i\bm{k}\cdot\bm{r}}/\sqrt{\Omega}$ with the system volume $\Omega$.  

A magnetic impurity with a toroidal moment scatters a plane-wave state from $\ket{\bm{k}'}$ to $\ket{\bm{k}}$.  Fermi's golden rule prescribes its scattering rate 
\begin{equation}
w_{\bm{k}\bm{k}'} = 2\pi |\braket{\bm{k}|\hat{H}_i|\bm{k}'}|^2 \delta(\epsilon_{\bm{k}}-\epsilon_{\bm{k}'}). 
\end{equation}
In the present model, the matrix element is given by 
\begin{align}
\braket{\bm{k}|\hat{H}_i|\bm{k}'} = \frac{1}{\Omega} 
\left[ V_i + \frac{\lambda}{2}\bm{T}_i\cdot(\bm{k}+\bm{k}') \right] e^{-i(\bm{k}-\bm{k}')\cdot\bm{r}_i}, 
\end{align}
where we assume short-range scattering $V(\bm{r}-\bm{r}_i) = V_i \delta(\bm{r}-\bm{r}_i)$ and $\bm{T}_i(\bm{r}-\bm{r}_i) = \bm{T}_i \delta(\bm{r}-\bm{r}_i)$.  Hence the scattering rate becomes
\begin{align}
w_{\bm{k}\bm{k}'} &= \frac{2\pi}{\Omega^2} \left[ V_i^2 + V_i \lambda \bm{T}_i\cdot(\bm{k}+\bm{k}') 
+ \frac{\lambda^2}{4} \left( \bm{T}_i\cdot(\bm{k}+\bm{k}') \right)^2 \right] \nonumber\\
&\quad \times \delta(\epsilon_{\bm{k}}-\epsilon_{\bm{k}'}). 
\end{align}

We define the symmetric and anti-symmetric parts of the scattering rate as \cite{Ishizuka}
\begin{align}
w^{+}_{\bm{k}\bm{k}'} &= \frac{1}{2} ( w_{\bm{k}\bm{k}'} + w_{-\bm{k},-\bm{k}'} ), \\
w^{-}_{\bm{k}\bm{k}'} &= \frac{1}{2} ( w_{\bm{k}\bm{k}'} - w_{-\bm{k},-\bm{k}'} ). 
\end{align} 
When time-reversal symmetry $\mathcal{T}$ is preserved while inversion symmetry $\mathcal{P}$ is broken, $w^{-}_{\bm{k}\bm{k}'}$ is equivalent to $w^\text{(A)}_{\bm{k}\bm{k}'} = ( w_{\bm{k}\bm{k}'} - w_{\bm{k}'\bm{k}} ) / 2$, which leads to skew scattering \cite{rectification}.
On the other hand, a toroidal moment breaks both $\mathcal{T}$ and $\mathcal{P}$ while the product $\mathcal{PT}$ is preserved.  
As a result, the scattering rate remains symmetric under the interchange of the initial and final states
\begin{equation}
w_{\bm{k}\bm{k}'} = w_{\bm{k}'\bm{k}},
\end{equation}
or $w^\text{(A)}_{\bm{k}\bm{k}'}=0$.  However, $w^{-}_{\bm{k}\bm{k}'}$ can be finite even in the first Born approximation.     
This is in sharp contrast to skew scattering, where the second Born approximation is essential to break the Hermitian nature of the scattering $T$-matrix \cite{Sinitsyn}. 
For the present model, we obtain
\begin{gather}
w^+_{\bm{k}\bm{k}'} = \frac{2\pi}{\Omega^2} \left[ V_i^2 + \frac{\lambda^2}{4} 
\left( \bm{T}_i\cdot(\bm{k}+\bm{k}') \right)^2 \right] \delta(\epsilon_{\bm{k}}-\epsilon_{\bm{k}'}), \\
w^-_{\bm{k}\bm{k}'} = \frac{2\pi}{\Omega^2} V_i \lambda \bm{T}_i\cdot(\bm{k}+\bm{k}') 
\delta(\epsilon_{\bm{k}}-\epsilon_{\bm{k}'}), 
\end{gather}
in the first Born approximation. 

We consider the case where $w^-$ is much smaller than $w^+$, namely $|V_i| \gg |\lambda T_i k_F|$, and treat $w^-$ perturbatively.  
The condition states that in Eq.~\eqref{eq:effective} the impurity potential is much larger than the spin-orbit term evaluated at the Fermi level. 
In calculating the response, instead of a single magnetic impurity with a toroidal moment as we have considered in the preceding discussion, we assume a dilute distribution of magnetic impurities each with a toroidal moment $\bm{T}_i$.  
In reality, the impurity distribution is not completely random, but impurities tend to occupy certain interstitial positions which are electrochemically stable.  It leads to a finite net toroidal moment when all impurity spins align in parallel.  
We denote the impurity density as $n_i$ and neglect interference of scattering at different impurities.  
Without loss of generality, we suppose that the toroidal moment points to the $z$ direction ($\bm{T}_i = T_i \hat{z}$).

Following Ref.~\cite{rectification}, we calculate the second-order response using the semiclassical Boltzmann transport theory.  We consider a three-dimensional system, but a generalization to other dimensions is straightforward.  
At low-frequencies $|\omega\tau| \ll 1$, we obtain the second-order conductivity tensor 
\begin{gather}
\chi_{abc} = \chi ( 2\delta_{az}\delta_{bc} + \delta_{ab}\delta_{cz} + \delta_{ac}\delta_{bz} ), \\
\chi = -\frac{8\pi e^3 \tau^3 n_i V_i \lambda T_i}{3m} \int d\epsilon [-f'(\epsilon)] \epsilon D(\epsilon).  
\end{gather}
$f(\epsilon)$ is the Fermi-Dirac distribution function and the transport scattering time $\tau$ is obtained from $w^+$ as 
\begin{equation}
\tau^{-1}
= \Omega \int \frac{d^3k'}{(2\pi)^3} w^+_{\bm{k}\bm{k}'} (1-\cos\theta_{\bm{k}'})
= \frac{1}{2} n_i V_i^2 D(\epsilon), 
\end{equation}
where $D(\epsilon)$ is the density of states of the conduction electrons. 
In the vector notation, we can describe the second-order current response in the general form 
\begin{align}
\bm{j}_2 = \chi [ 2\bm{t}_i(\bm{E}_1\cdot\bm{E}_2) + \bm{E}_1(\bm{t}_i\cdot\bm{E}_2) 
+ \bm{E}_2(\bm{t}_i\cdot\bm{E}_1) ], 
\end{align}
where $\bm{t}_i$ is the unit vector parallel to the toroidal moment ($\bm{T}_i = T_i \bm{t}_i$).  
This formula subsumes the nonlinear Hall effect \cite{NLH,NLH2} and the nonreciprocal magnetochiral anisotropy along the direction of $\bm{T}_i$ \cite{MA}. 

We have found that a toroidal moment induces second-order current response in the presence of the spin-orbit coupling.   
Even a single magnetic element can generate a toroidal moment when it is placed off an inversion center of the underlying crystal.  Our finding is thus applicable to a variety of systems, such as a metal with magnetic impurities and a thin film metal with magnetic deposits.
An external magnetic field controls the impurity spins and the toroidal moments, which modifies the second-order response and the nonreciprocal transport. 
When the direction of $\nabla V(\bm{r}-\bm{r}_i)$ is known, the direction and magnitude of $\bm{T}_i$ can be manipulated according to Eq.~\eqref{eq:T}. 
It vanishes when $\bm{s}_i$ is parallel/anti-parallel to the gradient of the impurity potential, while it becomes largest when perpendicular.  
Also, Curie's law is expected for a toroidal moment $\bm{T}_i \propto 1/T$ with temperature $T$ under a weak magnetic field, which provides further evidence of nonreciprocal transport by the toroidal scattering mechanism.

This work was supported by JST CREST Grant Number JPMJCR1874, Japan, and JSPS KAKENHI Grant number 18H03676.


\begin{thebibliography}{99}

\bibitem{Dubovik}
V. M. Dubovik and V. V. Tugushev, Phys. Rep. \textbf{187}, 145 (1990).  

\bibitem{Gorbatsevich}
A. A. Gorbatsevich and Yu. V. Kopaev, Ferroelectrics \textbf{161} 321 (1994).

\bibitem{Schmid}
H. Schmid, Ferroelectrics \textbf{252}, 41 (2001). 

\bibitem{Spaldin}
C. Ederer and N. A. Spaldin, Phys. Rev. B \textbf{76}, 214404 (2007).

\bibitem{MA}
Y. Tokura and N. Nagaosa, Nat. Commun. \textbf{9}, 3740 (2018).

\bibitem{Ishizuka}
H. Ishizuka and N. Nagaosa, Nat. Commun. \textbf{11}, 2986 (2020) 

\bibitem{rectification}
H. Isobe, S.-Y. Xu, and L. Fu, Sci. Adv. \textbf{6}, eaay2497 (2020). 

\bibitem{Sinitsyn}
N. A. Sinitsyn, A. H. MacDonald, T. Jungwirth, V. K. Dugaev, and J. Sinova, Phys. Rev. B \textbf{75}, 045315 (2007).

\bibitem{NLH}
I. Sodemann and L. Fu, Phys. Rev. Lett. \textbf{115}, 216806 (2015).

\bibitem{NLH2}
Q. Ma, S.-Y. Xu, H. Shen, D. MacNeill, V. Fatemi, T.-R. Chang, A. M. Mier Valdivia, S. Wu, Z. Du, C.-H. Hsu, S. Fang, Q. D. Gibson, K. Watanabe, T. Taniguchi, R. J. Cava, E. Kaxiras, H.-Z. Lu, H. Lin, L. Fu, N. Gedik, and P. Jarillo-Herrero, Nature \textbf{565}, 337 (2019)



\end{thebibliography}
\end{document}